# Entanglement swapping with photons generated on-demand by a quantum dot

F. Basso Basset[1,4], M. B. Rota[1,4], C. Schimpf[2,4], D. Tedeschi[1,4], K. D. Zeuner[3], S. F. Covre da Silva[2], M. Reindl[2], V. Zwiller[3], K. D. Jöns[3], A. Rastelli[2] and R. Trotta[1,*]

[1]*Department of Physics, Sapienza University of Rome, 00185 Rome, Italy*

[2]*Institute of Semiconductor and Solid State Physics, Johannes Kepler University, 4040 Linz, Austria*

[3]*Department of Applied Physics, Royal Institute of Technology, 106 91 Stockholm, Sweden*

[4]*These authors contributed equally.*

* rinaldo.trotta@uniroma1.it

**Photonic entanglement swapping, the procedure of entangling photons without any direct interaction, is a fundamental test of quantum mechanics[1] and an essential resource to the realization of quantum networks[2]. Probabilistic sources of non-classical light can be used for entanglement swapping, but quantum communication technologies with device-independent functionalities demand for push-button operation[3] that, in principle, can be implemented using single quantum emitters[4]. This, however, turned out to be an extraordinary challenge due to the stringent requirements on the efficiency and purity of generation of entangled states. Here we tackle this challenge and show that pairs of polarization-entangled photons generated on-demand by a GaAs quantum dot can be used to successfully demonstrate all-photonic entanglement swapping. Moreover, we develop a theoretical model that provides quantitative insight on the critical figures of merit for the performance of the swapping**

**procedure. This work shows that solid-state quantum emitters are mature for quantum networking and indicates a path for scaling up.**

Entanglement swapping has been observed in a few different systems, from the original all-photonic scheme which employs a spontaneous parametric down-conversion (SPDC) source[1] to hybrid protocols in which the interference of two photons is used to entangle spins[5] or atoms[6] at a distance. The swapping procedure between pairs of photons is especially relevant to the development of future quantum networks, because it provides a way to overcome the limitation against an optical communication amplifier for photonic qubits due to the no-cloning theorem and to create entanglement over distances beyond the reach of direct transmission[3,7].

Developing sources able to operate on-demand is an important step towards this goal. Despite the impressive technological achievements up to date[8], SPDC sources are limited by the probabilistic nature of the photon generation process, which can only be partially alleviated by heralding and active multiplexing[9]. Quantum emitters, such as atoms, nitrogen vacancies in diamonds and semiconductor quantum dots (QDs), overcome this hurdle and hold strong promise for deterministic operation. Among these, the latter are receiving attention after recent reports of QD-based single-photon sources overtaking SPDC in terms of brightness, single photon purity and indistinguishability[10,11], and they are closing the performance gap concerning the generation of polarization-entangled photons as well[12–14], leading to the recent demonstration of three-photon quantum teleportation[15], even under deterministic photon generation[16].

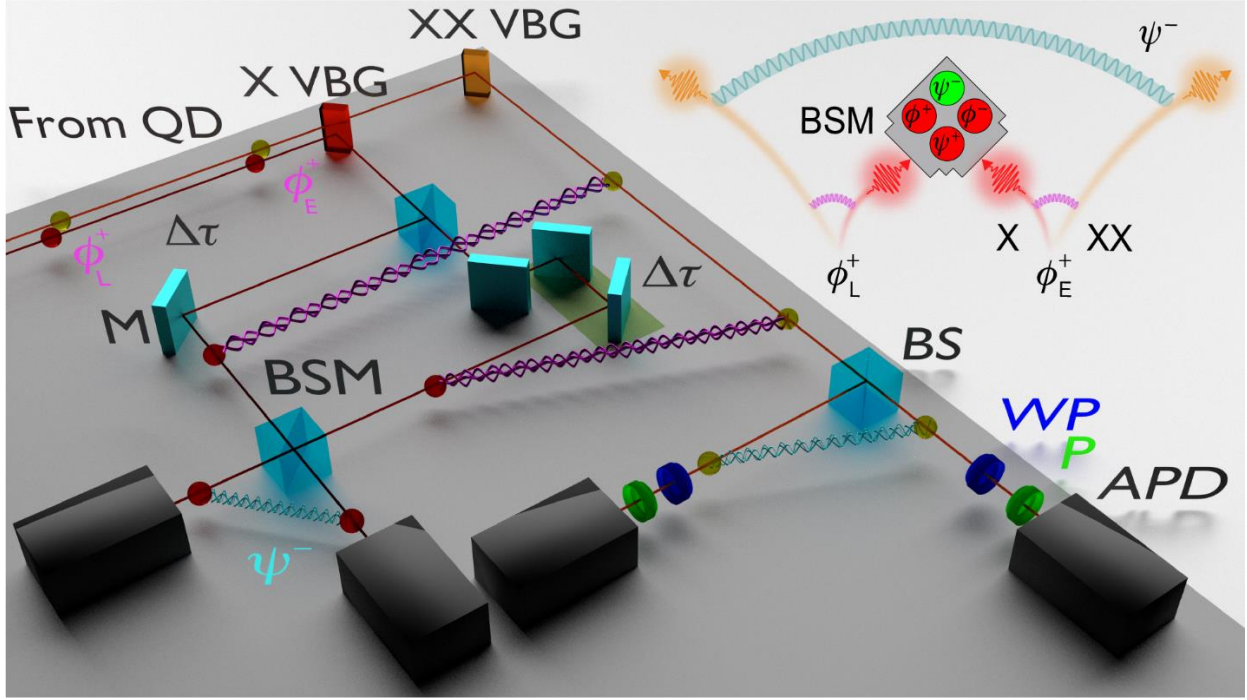

**Figure 1 | Experimental setup.** Schematic illustration of the setup used for the entanglement swapping experiment. XX and X photons are separated by volume Bragg gratings (VBG). X photons are sent into an unbalanced Mach–Zehnder interferometer with an internal delay Δτ matching the time distance between the early- (E) and late- (L) coming entangled pairs. The latter BS of the interferometer performs a partial Bell state measurement (BSM). XX photons are sent to a Hanbury Brown–Twiss analyzer composed of a non-polarizing beam splitter (BS) and two sets of wave plate (WP), linear polarizer (P) and single-photon counting avalanche photodiode (APD). In the inset, a sketch representing the ideal entanglement swapping process is shown.

To the best of our knowledge, however, there is at present no report on the use of QDs—and more in general of any solid-state-based quantum emitter—in all-photonic entanglement swapping schemes. This absence of experimental results is likely related to the outstanding challenges set by the implementation of four-photon swapping protocols involving high quality entangled-photon pairs. Below, we detail these challenges and explain how they can be successfully overcome.

Using a single QD we design an entanglement swapping experiment which follows the seminal work of Pan et al.[1], as illustrated in the inset of Fig. 1. Photon pairs in the maximally entangled state $|\Phi^+\rangle = 1/\sqrt{2}\,(|HH\rangle + |VV\rangle)$ are deterministically generated from a QD with exciton fine structure splitting (FSS) well below the radiative-limited excitonic linewidth[17] by exciting the radiative biexciton-exciton (XX-X) cascade via a two-photon resonant scheme[4]. Two XX-X entangled-pairs ($|\Phi_E^+\rangle$ and $|\Phi_L^+\rangle$) are independently triggered by two subsequent laser pulses, and the photons emitted by a specific transition ($X_E$ and $X_L$ in the case we choose the X transition) are brought to interfere at a beam splitter using a delay line. When these two photons are perfectly indistinguishable, a joint detection at the two output ports of the beam splitter corresponds to the detection of the state $|\Psi^-\rangle = 1/\sqrt{2}\,(|HV\rangle - |VH\rangle)$. This procedure, which implements a partial Bell state measurement (BSM)[18], allows the photons from the other transition ($XX_E$ and $XX_L$) to be projected onto the entangled polarization state $|\Psi^-\rangle$, despite being distant and previously uncorrelated.

This simple picture falls short when a realistic source is considered. Background light, multiphoton emission, finite FSS and possible decoherence processes can reduce the fidelity to the entangled state $|\Phi^+\rangle$[13]. Likewise, poor photon indistinguishability introduces mixedness in the output of the BSM that dramatically affects the swapping procedure. In addition, a bright source is essential because entanglement swapping relies on four-fold coincidence events whose rate has a steep fourth-power dependence on the light extraction efficiency (see Supplementary Section 5). Meeting all these requirements at the same time has proven to be challenging, precluding any implementation of quantum-emitter-based photonic entanglement swapping up to date.

However, no fundamental limit enforces the figures of merit of current entangled photon sources, and, in this work, we present a solution based on GaAs QDs grown by Al droplet etching[19]. Due to the fast radiative recombination and the limited exciton dephasing during the intermediate step of the cascade[14,20,21], a fidelity to a maximally entangled Bell state up to 98% has been reported using these emitters[13]. The use of a resonant two-photon excitation scheme also ensures on-demand operation with a fidelity of preparation of approximately 90%[4], excellent suppression of multiphoton emission[22] and good indistinguishability[20]. To improve light collection and enhance the brightness of the source, we integrate the QDs into a monolithic planar cavity composed of two asymmetric distributed Bragg reflectors. This convenient approach does not worsen the QD optical quality and guarantees a bandwidth large enough to deal with the wavelength difference between the X and XX transitions.

From such sample we select a QD with optimal trade-off among the relevant figures of merit (details about how the values listed below are measured are in the Methods and Supplementary Section 1). A low fine structure splitting (FSS) of 0.6(5) µeV and an exciton lifetime of 270(10) ps ensure a high degree of XX-X entanglement, as supported by the measured value of Bell state fidelity of 0.88(2). The Hong–Ou–Mandel (HOM) visibility of the X line is 0.63(2), due to background photons, beam splitter imperfections, and arguably to the time correlation within the radiative cascade[23,24]. Most importantly, a rate of approximately 0.5 MHz is achieved on the detectors recording the BSM, which results in a rate of four-fold coincidences of approximately 3 mHz, in agreement with our predictions on the throughput of the swapping protocol (see the Supplementary Section 5).

The reported count rates are measured in the setup sketched in Fig. 1. With a repetition rate of 160 MHz, a couple of laser pulses separated by a time delay of 1.8 ns excite the QD and trigger the emission of two pairs of entangled photons, which we label $XX_E$-$X_E$ and $XX_L$-$X_L$, linked to the early and late generation pulse. Volume Bragg gratings are used to separate the photons from the two transitions of the cascade, thus ensuring minimal losses. The photons originating from the exciton to ground state transition, $X_E$ and $X_L$, are sent to an unbalanced Mach–Zehnder interferometer (with an internal delay also set to 1.8 ns) to let them interfere at its second beam splitter. The BSM is thus performed by recording joint detection events between $X_E$ and $X_L$ within a time window of 0.6 ns. The photons from the biexciton to exciton transition, $XX_E$ and $XX_L$, are instead sent to a Hanbury Brown–Twiss analyzer for quantum state tomography. Only the detection events within a temporal range of 100 ns from a BSM are recorded.

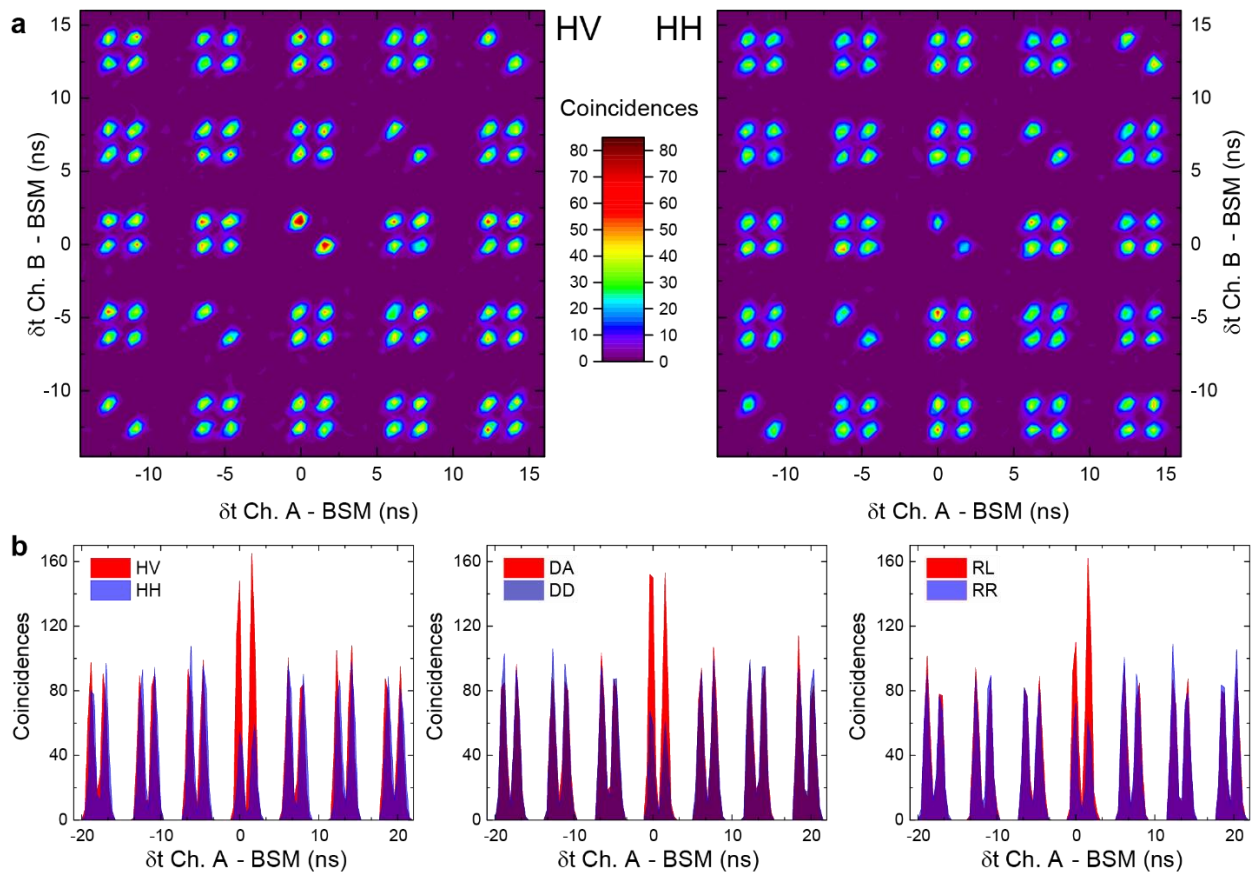

**Figure 2 | First evidences of entanglement swapping with photons from a QD. a,** Third-order intensity cross-correlation histograms among BSM and XX photons recorded for cross- (left) and co-polarized (right) linear polarization. Peaks along the main diagonal would correspond to XX photons excited from the same laser pulse and are therefore absent due to the single photon purity of the QD. The two peaks at the center belong to events synchronized with a BSM and differ only on whether $XX_E$ or $XX_L$ is detected on channel A. Bunching for HV and antibunching for HH are observed, as expected for the $|\psi^-\rangle$ state. **b,** Second-order intensity cross-correlation histograms between XX photons in linear, diagonal, and circular polarization bases. These data are reduced from third-order correlation histograms by binning over the time tags on channel B in the time window included between -1 and 2.8 ns.

Using the BSM as a trigger, we record third-order intensity correlation histograms, as shown in Fig. 2a for a pair of co- and cross-polarized XX bases. The comparison between the two peaks near zero delay—which contain the four-fold coincidences of photons coming from two subsequent XX-X cascades—highlights the presence of polarization correlation. To estimate the correlation visibility, the coincidence counts are normalized with respect to the side peaks due to XX photons uncorrelated with the BSM, as discussed more in depth in the Supplementary Section 2. In Fig. 2b the data are windowed and binned to obtain second-order intensity correlation histograms for the linear, diagonal, and circular bases. The observed bunching and antibunching behaviors clearly show the presence of a swapping process and are consistent with a projection to a state with a dominant $|\Psi^-\rangle$ character.

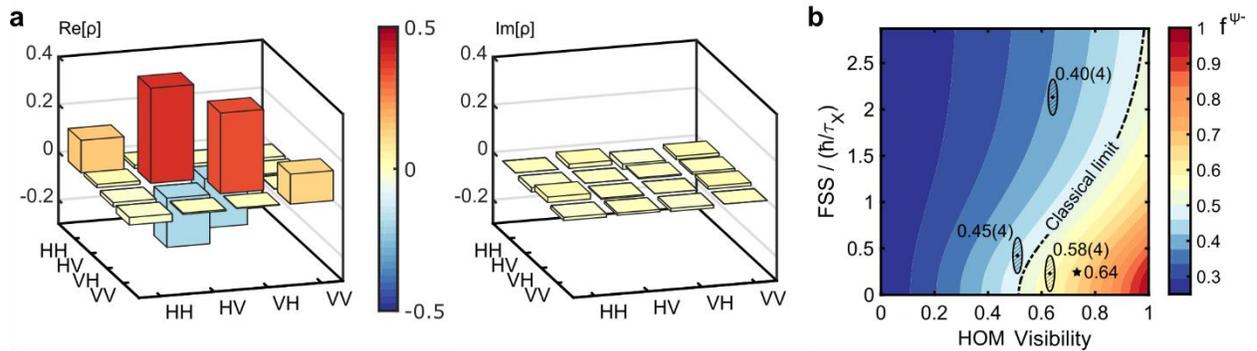

**Figure 3 | Density matrix of the XX photons after entanglement swapping. a,** Real and imaginary part of the two-photon density matrix reconstructed from measurement that describes the polarization state of the XX photons selected in conjunction with a BSM on their entangled partners. The matrix refers to the QD with FSS equal to 0.6 µeV and HOM visibility of 0.63. **b,** Contour plot showing the expected fidelity to the $|\psi^-\rangle$ state as a function of HOM visibility and the ratio between FSS and radiative lifetime. The ellipses refer to the QDs considered in this work, are labelled with the experimental values of fidelity to $|\psi^-\rangle$, and their semi-axes are given by the uncertainties on FSS and visibility. The star indicates the expected value of fidelity to $|\psi^-\rangle$ for the best QD in absence of background light and using a perfect beam splitter for the BSM.

In order to gain complete insight on the result of the swapping procedure, we perform the full tomography of the two-photon state[25] and collect $XX_E$-$XX_L$ correlations in the 36 possible combinations of linear, diagonal, and circular polarization bases[25]. Note that $XX_E$ and $XX_L$ are defined by their time of arrival and not by the detector that registers them, therefore permuted pairs of bases are acquired at the same time, and the total number of measurements is reduced to 21 (see Supplementary Section 3). The density matrix is reconstructed using a maximum likelihood estimation[26] and is presented in Fig. 3a.

The raw value of fidelity to the expected Bell state $|\Psi^-\rangle$ is calculated from the density matrix (see Supplementary Section 3) to be 0.58(4), which indicates a strong correlation between photons that did not interact, even surpassing the classical threshold[27] of 0.5 by two standard deviations. A consistent evidence of the presence of entanglement is offered by the above-zero raw value of the concurrence, equal to 0.15(8). Therefore, our results experimentally demonstrate entanglement swapping between single pairs of entangled photons generated on-demand by a source of non-classical light. In addition to that, the swapping procedure generates entangled pairs of photons with the same energy and different time bins. These are qualitatively different features with respect to cascaded photons usually observed in QDs.

It is worth emphasizing that the measured level of entanglement between the swapped photons does not consider non-idealities stemming from the experimental set-up, such as background light and non-ideal beam splitters. Taking these imperfections on board would push the fidelity to 0.64(4) and the concurrence to 0.28(8), see below. While this further strengthens our result, additional technical improvements are needed to reach the levels of entanglement needed for real-life quantum communication. To understand how to accomplish this task, we develop a theoretical model that not only accounts for the experimental observations, but it also pinpoints the next steps ahead.

The success of entanglement swapping critically depends on two main parameters: the initial degree of entanglement of the photon pairs and the indistinguishability of the photons involved in the BSM. The initial degree of entanglement is known to be limited by finite FSS, spurious photons from background light or multiphoton emission, and decoherence mechanisms during the intermediate step of the cascade[13]. From a theoretical point of view, it is possible[28] to introduce all these contributions in the density matrices of the initial $|\Phi_E^+\rangle$ and $|\Phi_L^+\rangle$ states (that are $\rho_{X_E,XX_E}$ and $\rho_{X_L,XX_L}$) and, after projecting the two X polarization modes onto $|\Psi^-\rangle$[29] and integrating over the exciton recombination times, in the swapped density matrix $\rho_{XX_E,XX_L}^{\psi-}$. However, a value of indistinguishability between $X_E$ and $X_L$ below unity reduces the probability that a joint measurement at the two outputs of the beam splitter accurately heralds $|\Psi^-\rangle$. We take this effect into account by calculating $\rho_{XX_E,XX_L}^{\psi-}$ as a weighted sum over the possible outcomes of the BSM. Eventually (see the Supplementary Section 6), we can derive an analytic expression for the fidelity to $|\Psi^-\rangle$, which reads

$$f_{XX_E,XX_L}^{\psi-} = \frac{1}{4}\left(1 + \frac{V}{2-V}k^2\left(g_{H,V}^{\prime(1)^2} + 2\frac{g_{H,V}^{(1)^2}}{1+\left(\frac{S\tau_X}{\hbar}g_{H,V}^{(1)}\right)^2}\frac{1}{1+\left(\frac{S\tau_X}{\hbar}g_{deph}^{(1)}\right)^2}\right)\right) \quad (1)$$

where $k$ is the fraction of uncorrelated photons collected from the XX-X cascade, $S$ is the FSS, $g_{H,V}^{\prime(1)} = 1/(1+\tau_X/\tau_{SS})$, $g_{H,V}^{(1)} = 1/(1+\tau_X/\tau_{SS}+\tau_X/\tau_{HV})$, and $g_{deph}^{(1)} = 1/(1+2\tau_X/T_2^*)$ with $\tau_{SS}, \tau_{HV}$, and $T_2^*$ respectively defined as the spin-scattering, cross-dephasing, and pure dephasing characteristic times, and $V$ is the HOM visibility.

By measuring $g^{(2)}_{X,XX}(0)$, $\rho_{X_E,XX_E}$, $\tau_X$, and $V$, and taking the value of the decoherence times from the literature[13,28,30] it is possible to experimentally estimate all the quantities appearing in Eq. 1 and, therefore, predict $f^{\psi-}_{XX_E,XX_L}$ with no fitting parameters. The model returns a swapping fidelity of 0.56 (0.64 in absence of background light and considering beam splitter imperfections), in excellent agreement with the experimental result. As a further proof of our theoretical model, we repeat the experiment and intentionally decrease either the degree of entanglement of our source, selecting a QD with a larger FSS of 5.9(5) µeV, or the indistinguishability of the photons, using an emitter with a HOM visibility of 0.51(2). The comparison between these data and the model, summarized in Fig. 3b and discussed in more detail in the Supplementary Section 6, shows once again good agreement and quantitatively indicates how the outcome of the swapping depends on the most critical properties of our quantum emitters, i.e. photon indistinguishability and initial degree of entanglement.

This quantitative agreement between experiments and theory suggests the steps needed to achieve a near-unity degree of entanglement between the XX photons after the swapping procedure. A greater effort should be devoted to improving the photon indistinguishability, which is directly affected by phonon-induced dephasing as well as by time correlation in the XX-X cascade. Improvements can be achieved using broad-band cavities enabling and independently tailoring Purcell enhancement for the X and XX transitions[31]. The use of photonic cavities will also allow boosting the flux of QD photons, a must when real-life applications are to be considered[32]. On the other hand, the degree of entanglement of the XX-X cascade can be pushed to near-unity values via multiaxial strain-tuning[33], a technique which is already available for integration and that enables interfacing QD photons with other quantum systems[34]. Considering the enormous progress witnessed in the last few years in the field of solid-state quantum photonics, we envisage that these tasks will be accomplished in the near future, and our demonstration of entanglement swapping opens the way towards the realization of a quantum network involving deterministic sources of entangled photons.

**Methods**

**Entangled photon source.** The sources of polarization entangled photon pairs are single QDs made of GaAs in a crystalline matrix of $Al_{0.4}Ga_{0.6}As$. The QDs were fabricated on a GaAs (001) substrate in a molecular beam epitaxy system at JKU Linz using the Al droplet etching technique[35]. The growth parameters were controlled to yield nanostructures with high in-plane symmetry[19]. The QDs were placed at the middle of a 123 nm-thick layer of $Al_{0.4}Ga_{0.6}As$ inserted between two 60 nm-thick layers of $Al_{0.2}Ga_{0.8}As$, which together constitute a λ-cavity. Below and above this cavity we placed distributed Bragg reflectors made of 9 and 2 layer pairs respectively, each composed of 70 nm of $Al_{0.95}Ga_{0.05}As$ and 60 nm of $Al_{0.2}Ga_{0.8}As$. The sample was finally capped by 4 nm of GaAs to prevent oxidation. Even if no Purcell enhancement is observed, the collection efficiency is greatly enhanced by the insertion of the distributed Bragg reflectors. Including the effect of a half ball lens made of N-LASF9, an extraction efficiency of approximately 10% is estimated in the spectral region near 785 nm.

**Entanglement swapping setup.** During the experiments, performed at Sapienza Rome, the QDs are kept at a temperature of 5 K in a low-vibration closed-cycle He cryostat from attocube systems. They are excited with a resonant two-photon excitation scheme[4] using a Ti:Sapphire femtosecond laser with 80 MHz repetition rate. An unbalanced Mach–Zehnder interferometer with a delay of 6.25 ns allows the effective repetition rate of laser pulses to be doubled, while a second interferometer with 1.8 ns delay is used to prepare for synchronous photon detection in the BSM and HOM setup. The spectrally broad laser pulse is sent through a 4f pulse slicer to reduce its linewidth down to approximately 200 µeV.

The laser is focused onto the sample by a 0.81 NA objective, placed inside the cryostat. The objective also collects the signal. The laser backscattered light is filtered out by tunable volume Bragg gratings with a bandwidth of 0.4 nm. A second set of volume Bragg gratings used in reflection mode spectrally separates the emission from the two transitions of the XX-X cascade, as shown in Fig. 1a. The BSM and the two-photon interference are performed using a single-mode fiber-coupled beam splitter with 48% reflectance, 52% transmittance and a mode overlap of 96%. The Hanbury Brown–Twiss analyzer for quantum tomography is composed of a nonpolarizing beam splitter followed by zero-order quarter- and half-wave plates and nanoparticle linear film polarizers. Polarization controllers are used to compensate for unitary transformations of the polarization state of X and XX photons induced by the setup, as described more in detail in Supplementary Section 4. During the correlation measurements, the signal is detected by four silicon avalanche photodiodes with a time jitter of approximately 400 ps and a quantum detection efficiency of approximately 65%. The single photon counts are recorded by time tagging electronics with 10 ps resolution and analyzed on the fly to yield the third-order intensity correlation histogram among the BSM coincidences and the detections on the XX channels.

The same setup is used to measure the autocorrelation function of the XX and X emission, the XX-X entanglement and the HOM visibility.

A faster silicon avalanche photodiode with time jitter slightly above 50 ps is used for lifetime measurements. QDs emission spectra are acquired by a deep-depletion, back-illuminated LN2-CCD camera using a 750 mm focal length spectrometer, equipped with a 1800 g/mm grating. The FSS is estimated from polarization-resolved spectra acquired by adding a rotating half-wave plate and a linear polarizer to the collection path[36], resulting in an accuracy of approximately 0.5 µeV.

**Error analysis.** A Monte Carlo approach is used to estimate the error of the physical quantities calculated from the density matrix. For each correlation measurement possible data inputs are randomly generated from a Poissonian distribution whose average is the measured value of coincidence counts. This procedure is iterated 2000 times to obtain a large enough sampling for consistent results. The density matrix, together with the related physical quantities of interest, namely the Bell-state fidelity and the concurrence, is calculated for each set of simulated entries. The standard deviation of the obtained outcomes gives our error estimate.

**Acknowledgements**

This work was financially supported by the European Research Council (ERC) under the European Union's Horizon 2020 Research and Innovation Programme (SPQRel, grant agreement no. 679183), the Austrian Science Fund (FWF; P 29603), and the European Union Seventh Framework Programme (FP7/2007-2013) under grant agreement no. 601126 (HANAS). K.D.J. and R.T. acknowledge the COST Action MP1403, supported by COST (European Cooperation in Science and Technology). A.R. acknowledges the Linz Institute of Technology for support and Y. Huo and G. Weihs for fruitful discussions. K.D.Z. gratefully acknowledges funding by the Dr. Isolde Dietrich foundation.

**Authors' contributions**

F.B.B., M.B.R., C.S., and D.T. performed measurements and data analysis with the help of K.Z., K.D.J. and R.T.. S.F.C.d.S., and A.R. designed and grew the sample. F.B.B., M.B.R., C.S., D.T. and R.T. wrote the manuscript with feedback from K.Z., M.R., K.D.J., V.Z., and A.R.. All the authors participated in the discussion of the results. R.T. conceived the experiments and coordinated the project.

**Competing interests**

The authors declare no competing financial interests.